*Research Article*

# Pareto-Nash Reversion Strategies: Three-Period Dynamic Co-operative Signalling with Sticky Efficiency Wages

Mayaki, Alfred A. B.

Institutional Affiliation: The Open University (UK)
Contact E-mail: alfred.mayaki@ou.ac.uk

Abstract

This paper uses Nash equilibrium reversion as an optimal tool for clearing dynamic prices and wages. Various exogenous competitive rigidities determine the balanced growth path of the efficiency wage and the outcome of repeated household/firm wage bargaining decisions. A location model is used to explore the extent to which a downstream spatial cooperation agreement might affect the price equilibrium. There is also an endogenous hiring function and a knowledge base that is increasing in output as is the real wage. As the article demonstrates after accounting for real rigidities in the baseline model the effect of wage growth on household utility through staggered bargaining can be best captured by adopting a policy of point scoring on the mobility of skilled labor against the model's key rigidities. Mobility point scores which serve to encourage mobility from skilled labor within the model not only increase the knowledge base but also place upward pressure on nominal wage growth.

Keywords: Labor, Signalling, Mobility, Wages, Game Theory, Competition

JEL Classification: D19, D21, J31, J61, L16

## 1. Introduction

Research has progressed over the last 20 or more years, and models have been established that have unquestionably furthered the understanding of equilibrium labor conditions in both British and American economies (Pissarides, 1988). There have been several key papers written on the importance of the 'matching function' (Mortensen and Pissarides, 1999; Blanchard, 1991; Coles and Smith, 1996), as well as models of wage determination (Blanchard, 1982), including but not exclusive to noncooperative bargaining games (Yellen; 1984).

There is also an extended generation of contemporary research on the relationship found between prices and wages in cross-sectional empirical data. However, there has been relatively little attempt to explain a model of *conglomerate* price competition and wage behavior throughout the full business cycle prior to and after an output shock.

### *1.1 Background*

Economic growth over the last 20 years has occurred at a remarkable rate, particularly in developed economies. The acclaim for such successes should no doubt be placed on the heads and shoulders of the increasing aptitude of the three or four overlapping generations who have joined and retired from the British and American labor markets and who will continue to do so. The actual number of these participants has increased, but not without leaving behind adequate developments in the underlying tools of their and all our productive capabilities, i.e., *technology*. For good reasons, this article emphasizes the point that such growth may very well continue or may not. Whichever scenario occurs over the course of the next century, I would have it that developed economies, with time, must enshroud their efforts to encourage the mobility of skilled labor into their domestic markets, or as Romer (2012) terms the effect, such economies must discover 'the contribution of a mystery variable, "*the effectiveness of labor"*, whose exact meaning is not specified by Romer and whose behavior is taken as exogenous (independent of explanatory variables), although it is alluded to the paradigm of research and development.

As such, welfare should be discouraged, necessarily because possible margins for increases in per capita output can emerge from the productivity of the labor market and a reduction in government expenditure on social security. Effective labor and the value of the knowledge economy must be prioritized. In the following baseline model, this variable equates to none other than the level of household '*effort'*, derived from a combination of 1) added work value and 2) the value of qualitative contributions – concepts that are all too often over-assumed in classical economic growth theory. Of course, as luck would have it, the exact relevance of this variable is its relation to the essence of this article, the mobility of the worker (or rather, the *lack* of it) and their effective level of remuneration.

Many authors have taken steps to attempt to explain the irregularity created by sticky efficiency wages either by constructing new models that build on Keynesian theories concerning wage growth (Summers, 1988; Akerlof and Yellen, 1985; Hall, 2005) or by building on newer more dynamic theories. Rapid wage growth (from $w\ to\ w*$) is a staple of the following aggregate baseline model and, as we will come to see, depends somewhat partially on the ubiquity of skilled labor inputs. Due to the nature of dynamic stochastic general equilibrium (DSGE) models, there are only a finite number of identical households and firms, each of whom is assumed to have zero switching (or menu costs) associated with job separation, destruction and creation.

*1.2 Materials and Methods*

This thesis begins by establishing a framework based on the Pissarides/Mortensen matching function, which allocates values to each determinant of an equilibrium real wage, resulting in a broad and encompassing definition of the labor market, its participants, and the tools with which they negotiate. The model in the article varies from that of Pissarides/Mortensen in terms of the proposition of a hiring rate, which replaces the conditions for job creation and destruction. The next part of the article prepares the conditions for analyzing wage behavior and firm competition by setting out the conditions for a general Nash wage bargaining solution.

## 2. Literature Review

*2.1 Conglomerate Pricing: Entry Signalling*

Take, for example, the case of a firm that wishes to optimize output and a real wage in a signalling and bargaining scenario, i.e., a period-by-period staggered wage and an infinitely repeated Bertrand oligopoly game, where an incumbent's profit function is diversified across industries. What is the central question? This article seeks to present a theoretical study of the effect of *entry* signalling on the behavior of an oligopoly, which includes at least one subsidiary firm (i.e., a 'conglomerate of firms') and the wage decisions such firms make across time, i.e., dynamically. Whereas much of the previous oligopoly literature has focused on three or more single-segment firms that compete based on price, little work has been published on the effect of price competition on a corporate group or affiliate of firms operating under the same 'body corporate' and wage minimization/profit maximization decisions that such firms undertake. In such a context, collusion can often be a factor that determines expected future profitability. Observing the coordinated correspondence between two or more conglomerate firms and a market entrant presents a unique opportunity to achieve greater social output as well as optimal profit given total net demand.

*2.2 Equilibrium conditions*

By converting a Bertrand-Nash equilibrium price into a cooperative equilibrium, the cooperative subgame perfect wage equilibrium becomes a function of the optimal Pareto-efficient *argmin–max* payoff. The seminal papers on oligopoly theory (Friedman, 1971; Harrington, 1991; Abreu, 1986; 1988) all establish a key element of this article, that is, the modeling of some sort of cooperation prior to simultaneous (and often secretive) profit-maximizing Nash reversion (at the expense of entry signalling). Thereafter, each firm may choose to undercut and converge to a one-shot Nash equilibrium for the remainder of the infinite game. The obvious concern with constructing barriers to entry is that Nash reversion creates significant increases and entry gaps that are likely to inflate discounted prices and wages and generally reduce consumer and social welfare.

More recently, however, the noneconometric focused research literature on this topic has become increasingly important, with an emphasis on optimization in coordinated equilibria across markets and within a variety of environments (Dal Bo and Frechette, 2011; Rees, 1993), chiefly because each oligopolistic firm has to signal entry into an upstream market and negotiate with new laborers.

*2.3 Shirking and Staggered Nash Wages*

Much of the seminal work on wage behavior found in Yellen (1984), Akerlof and Yellen (1985), Stiglitz (1976), Blanchard and Katz (1999), Mortensen and Pissarides (1999) and Diamond (1996) refers to the concept that the *nominal* wage has been used to explain, e.g., the absence of wage depreciation during aggregate shocks. This article first highlights that the 'shirking' model of Shapiro and Stiglitz (1984) is essential for understanding the efficiency wage during recessions. The intuition that firms pay above market wages, i.e., an *efficient* wage to ensure that their workers do not shirk under contract, while maintaining an equilibrium price is conventional wisdom. However, firms may agree that this equilibrium wage is indispensable for retaining their skilled labor and optimizing their marginal product, in which case the added incentive used to deter shirking disappears. The price/quantity duality will always remain constant. The unintended consequence is thus underemployment (for the agent who strictly prefers shirking to effort regarding wages).

*2.4 Augmenting the G-T Model*

A robust DSGE model of staggered multiperiod wage contracting (Gertler and Trigari, 2009) is the first point of reference. G-T is important for two main reasons. First, we are trying to understand the wage effects on household utility and firm output during entry games, where households and firms renegotiate wages based on a Nash perfect subgame through the use of variations in hiring by firms, which is justified on the basis of studies by Hall (2005) and Shimer (2005), among others. The work of the G-T model, unique in that it provides intuition that precedes the financial crisis, is valuable because the understanding of the model concerning *wage* stickiness prior to an actual fluctuation in economic output vis-à-vis.

Second, the G-T model associates wages with hiring, where the stickiness of the wage due to staggered contract negotiations implies that the current and expected movement in the marginal product of labor ($L$) and capital ($K$) will have a greater impact on the hiring rate ($h$) than would have been the case otherwise. The objection here is that economists such as Barnichon (2009) have studied wage effects on the Mortensen and Pissarides (1999) matching function with more recent models and can testify to incorporating cyclical fluctuations with job separations, using even more recent work from Shimer.

## 3. Households and Workers

The model population consists of $H$ finite households. The population grows at a constant rate ($g$) due to the contribution of incoming workers who supply skilled labor ($L$s). The average household's problem is to choose the exact path of the efficiency wage $w(t)$ to maximize the present value of lifetime utility, given the following optimization problem. where $A$ represents knowledge endowments, $e$ represents effort, e.g., versus time spent on leisure, and $n$ represents households that can form profitable alliances. By providing the integral with preferences, we obtain

*3.1 Basic Utility Function*

$$U = \alpha + e(n, A, H, L, w)g + dt$$

The variable $\alpha$ represents a positive discrete constant integer that equates to initial household wealth. The higher this value is, the less intrinsic incentive the worker has to work. The expression for labor ($L$) is divided by the time supplied to the labor market against leisure at time $t - 1$ by $H$ households, where $A$ is the level of productivity (effective knowledge), which is given as exogenous.

The growth rate of $A$ and the growth rate of $g(t)$, which is the growth rate of the population, do not necessarily increase the value of the household's instantaneous utility at time $t = 0$ as the function increases in $w$.

*3.2 Budget Constraint*

Take the following evolution of the previous utility function:

$$\alpha + \frac{L(t)}{w(t)} g(t) + A(t)$$

Expanding this equation by adding the optimal level of labor for the economy, that is, $L$ (hat) less $L$ at time $t$. The integral for technological '*knowledge*' growth is taken as given.

$$U = \int_t^T \sum e^{-R(t)}\, \alpha(t) + \frac{(\bar{L} - L)(t)}{w(t)H(t)} g(t) + \int_t^T A(t)\, dt$$

$$s.t.$$

$$\alpha - (1 - Ln)(e)g$$

Taking the first-order conditions of the household's instantaneous utility function, we derive an expression that reflects the budget constraint as a function of utility:

$$U' = \int_t^T \alpha'(t) + L'(t)g'(t) + \int_t^T A'(t)\, dt$$

When the constraint is binding, total workforce effort is equal to the integral of the discounted value of future lifetime earnings of the present population. The earnings of generation $t + 1$ are given by the remaining endowments left at the end of time $t = 0$ less fiscal responsibilities carried over to the next generation. The household therefore wishes to maximize total utility by trading off its perceived leisure time with its ability to offer labor as a member of the workforce and by dissaving its initial wealth endowment α.

## 4. Competitive Firms

Given that Walrasian assumptions hold for firm $f$, who wishes to hire a marginal unit of labor $L^m$, offering labor to a firm $f$ are $H/L$ number of workers, infinitely lived, who are invariably in and out of the market for their labor each competing for a finite number of homogenous jobs $J$. Each firm faces the following Lagrangian optimization problem:

$$Q = p(1-q)^2 \qquad s.t.\ \ px(1-q) + wl \leq w + M$$
$$s.t.\ \ l \leq 1$$

*4.1 Cooperative Games*

The effect of cooperative behavior prior to a simultaneous (and often secretive) profit-maximizing Bertrand Nash reversion is the expense of lower barriers to entry. A co-operative subgame perfect equilibrium of an infinitely repeated Bertrand oligopoly game is a function of the optimal Pareto-efficient payoff. This is opposed to the other theories on Cournot–Nash reversions (Abreu and Pearce, 1986) or Stackelberg competition (Green and Porter, 1984). Thereafter, an entry game converges from a dynamic equilibrium to an infinite 1-shot Nash equilibrium (Baye and Morgan, 1999). More recently, more non-econometric-based research on Nash reversions has become increasingly important, with an emphasis on optimization in games with co-ordinated deviation within a variety of environments (Dal Bo and Frechette, 2011; Rees, 1993), chiefly due to increasing dynamic incentives.

To a large extent, microfoundational research has focused on two main forms of punishment strategy: Abreu-type punishment strategies (Abreu, 1988) and the Nash-reversion (Friedman, 1971) or the *grim-reaper* strategy. Recent studies have demonstrated the effect of punishment strategies on the outcome of repeated games (Wright, 2011), as well as a welcome to be effective in using case study examples to demonstrate how concentrated markets are most effective in forming collusive outcomes (Zhang and Round, 2011).

*4.2 Competive Entry and Profit Maximization*

Take, for instance, the Matutes and Regibeau (MR) two-stage model, where submarket participation is not considered in stage one, where decisions are subsequently made regarding product diversification or homogeneity standardization. It is assumed that wage determination is also achieved. The conglomerate must then decide upon the use of a Pareto multistrategy $m \in M$, competing as part of two subsidiary markets, one accompanied by standardized goods and a constant price $P^m$, and the other with nonstandardized goods and more severe competition. The entrant pays no fixed fee (Matutes and Regibeau, 1989:361). By solving the unique subgame perfect equilibrium $\tilde{\pi}$, for the sequential M-R model, which is derived through backward induction, where the demand for each submarket ξ is equal to: By simultaneously undercutting the incumbent, the competition concentration ratio increases, an entrant Pareto-optimal profit function is given by the following: profit from undercutting the incumbent less profit gained by colluding $\gamma - c$.

If $\gamma > c$ for all $p^{\Omega} > p*$, $e^i$ then entry occurs if an entrant can control for the competition created by $p^{\Omega} - pi$ via n $e^n$ number of entrants through limit pricing strategies. However, as the number of entrants increases, perfectly competitive profits converge to zero. The likelihood of a collusive outcome in a cooperative oligopoly market is much greater because the incentive to revert is minimized (Zhang and Round, 2011:364).

*4.3 Standadisation Across Markets*

Matutes and Regibeau (1989) claim that consumers bundle goods from different firms to form a coherent 'system', creating incentives for all firms that offer a part of such a system to offer as close to an undifferentiated good as possible (when the consumer has already acquired the majority of their system). The incumbent firm, pursuing entry signals, faces an optimal strategy choice, either to sell a differentiated product and risk losing consumers due to incompatibility with what may be considered to be market niche 'systems' or to impose a uniformity constraint on the degree of diversification and prioritize price competition across markets. Firms may prefer incompatibility to distinguish customer incentives from competition, tying them in. Industry demand is increased when every firm produces all the component parts of a system, ensuring that they are each undifferentiated in their own respect. The incentive to undercut is reduced, as is the incentive to compete.

*4.4 The Pissarides/Mortensen Endogenous Hiring Rate*

The hiring function is preceded by the following intuition. There are an infinite number of identical homogeneous firms (*f*) that simultaneously utilize labor inputs to increase their respective productive capacities. Offering labor to these firms includes 1) an infinite number of households (*H*) that are infinitely lived and invariably in and out of the market for their labor, each competing for a finite number of homogeneous jobs given by $(j)^n$. At any given time (*t*), there are consequently ($e^{mc}$) numbers of employed households and ($e^{u}$) numbers of unemployed households. This constitutes the labor market. The model takes the endogenous rate at which a firm's hire labor (*h*) is at a constant between $-1 < h < 1$. In a perfectly competitive economy, these conditions are given through the *marginal* revenue product of labor for each firm at the given period of time *t*, interpreted by:

$$x \;=\; (\Delta\pi \;/\; \Delta e^{m})t$$

Firms are incentivized to participate in labor capital and markets until the marginal product reaches a reservation productivity threshold, at which each firm sets accordingly. Therefore, firms willingly and invariably hire households at the hiring rate $h$. Firms can now post vacancies, if and only if

$$x \;=\; \left({}^{\Delta\pi}/_{\Delta e^{m}}\right)^{t} + (\pi \;/\; e^{m})^{t-n} \geq 0$$

This generally implies that the number of vacancies at any point in time t is set by an aggregate hiring rate, given that the inequality constraint is binding. The constraint function is given by:

$$x \;=\; (\pi \;/\; e^{m}\;)^{\mathrm{t-n}}$$

which takes the average marginal product of the firm from t-n previous periods. This provides an accurate reflection of optimal labor input given the firm's revenue, the point at which firms reach labor demand/supply equilibrium and are unwilling to change labor inputs. An important feature of the model is the following:

$$h \;\in\; (0,1) \;\rightarrow\; 0, + (\Delta\pi \;/\; \Delta e^{m}\;)^{\mathrm{t}} \rightarrow (\pi \;/\; e^{m}\;)^{\mathrm{t-n}}$$

That is, as the aggregate hiring rate approaches positive (signed) zero, firms inevitably choose to reduce the number of vacancies they post to the market to satisfy reservation productivity. Furthermore, if $h$ is an extremely low figure (e.g., 0.01), firms have either reached, or are extremely close to reaching, an optimal output. The lower $h$ is, the greater the unemployment. Hence, the rate is also an avid indicator of structural employment. For any reason,

$$(\;\pi \;/ e^{m})^{t-n} \;>(\;\Delta\pi \;/\Delta e^{m})^{t}$$

That is, reservation productivity outweighs the value of weighted productivity; all things being even, excess inefficiency will lead firms to take action to reduce the period wage. That is, the change in MRPL is relatively low in comparison to the change in the real wage from the previous period. This will become the central theme of the discussion of wage bargaining. For a firm without a binding inequality constraint, where the MRPL is less than the reservation productivity, reducing the efficiency wage for a period has an identical effect, similar to that of job destruction. As mentioned before, job creation (creating vacancies) in the Pissarides model occurs at every point up until the optimal MRPL is reached, after which the hiring rate becomes negative, where h ∈ (−1, 0), the conditions are in place for job destruction (reducing $e^{m}$ and increasing $e^{u}$). Hence, on a very general note, the sticky efficiency wage ($w$) in this process implicitly facilitates the adjustment to a new equilibrium.

Finally, firms engage in implicit or explicit behavior both in the consumer market and in the labor market aimed at reducing the MRPL to signal entry into a conglomerate/oligopolistic market and to reach an optimal output implicitly by adjusting the real marginal wage offered under a Nash bargaining game or explicitly through job destruction or $h < 0$ negative hiring, i.e., reducing the total number of posted job vacancies (and consequently increasing outflows into unemployment).

*4.5 Staggered Wage Bargaining*

A closed-form staggered wage bargaining solution (Cheron and Langot, 2004; Gertler and Trigari, 2009) is used to understand the growth of the sticky efficiency wage. Contractual agreements take place once individual preferences have been established to optimize joint (worker and firm) utility. In other words, the wage of the new worker who enters the market for labor is the result of a co-operative Nash reversion between the firm (*f*) and the new worker *c*. The new worker has strict preferences that are complete, transitive and reflective. The new worker also has utility preference *X* such that $u$: $X \rightarrow R$ where $x > y$. Firms have utility preference *Y* if they agree to a high wage. The payoffs are reversed if they agree on a lesser wage. If they cannot agree to enter a contract, both parties end up with zero utility. Wage negotiation takes the form of a period game. In each period, wages are renegotiated by the firm and its new workers as a function of firm output, price level, capital and labor. We are interested in the subgame perfect equilibrium. We assume that there is perfect monitoring, i.e., that the past actions of all the firms are known. All assumptions pertaining to the histories and pure strategies of repeated games are taken as given, i.e., monotonicity, nonsatiation, convexity and continuity.

$$(4.5.1) \quad \pi_i = (p1, p2, \dots, pn) \;=\; \begin{cases} \pi_{(pi)} & \text{if } P_1 < \; P_t \\ \pi_{(pi)} & \text{if } P_1 = P_b \\ 0 & \text{if} \quad \text{otherwise} \end{cases}$$

Where there are past deviations from the Nash reversion, (i.e., if the firm chooses to rely on its agreement to pay new workers an amount that is equal to or strictly greater than the previous period wage, deviations are punished in the future by less effort and productivity). Future payoffs (*wages*) can be discounted (to the *present* period of time) with a discount factor. Are there any drawbacks to an approach that calculates the minimal payoff discount factor under which collusion can take place? Perhaps, for instance, calculating the net present value of future lifetime earnings is a problem for satisfying the budget constraint of the household maximization problem stated earlier.

Let us start with the assumption that firm $f$ then chooses to undercut other firms by negotiating a lower wage with new worker $c$ (nash bargaining reversion). This downward shifting effect that new workers have on an economy after average firm output has been reduced is essentially sticky wages. The outcome of the Nash reversion is simple. The model explains two intermittent utility functions that meet at equilibrium, one that is quasiconcave and increasing in $w$ and the other that must be quasiconvex and decreasing in $w$.

Firms and workers discuss the possibility of entering into a contract for employment. We can calculate the simple Nash equilibria of the game by undertaking a systematic analysis of the best responses to various strategic choices by each player.

## 5. Results

When hiring, each firm considers the marginal revenue product of labor, seeking to find $argmax(\Delta\pi)$, a function that maximizes revenue, given wage and inventory costs. What exactly is the relevance here? This means that there must be a saddle point with respect to the linear profit and cost functions, where the revenue maximizing output is set to $argmax(\Delta\pi) = argmin(w)$. On the other hand, $H$ households can only bargain individually to maximize their wage $argmax(w)$. Therefore, for each observation on the Beveridge curve, the market wage can be said to be a process of Nash bargaining where the households $argmax(w)$ equals the firms $argmin(w)$. Therefore, from this, it is possible to state that movements along the curve are characterized by changes in endogenous variables under the control of households and firms, which are in turn subject to a series of wider exogenous factors. Before and after the steady-state, the real wage for the labor market is given by the Nash solution, the outcome of $f^n(e)$, the number of $argmax(w)e = argmin(w)f$ ultimatum games between each unemployed worker € and each of the corresponding firms (*f*) on the basis of wages $w$. The bargaining outcome may be two kinds, either agreements or disagreements (often called disagreement points). Disagreements are typically defined by (Z) where $Z = (ze, zf)$, and agreements are noted by (*W*) where $W = (we, wf)$. The Nash solution function describes the process as $n$ number of $f$ firms (who post vacancies) consider maximizing labor productivity at a given output $q$ and $n$ number of $e$ unemployed households bargain for each vacancy, considering the real inflation-adjusted market wage offered by the firm against the wage being paid to employed households. The bargaining solution can be sketched in extensive form.

Here, we can assume that if the co-operative optimal payoff is not affordable for the firm to agree with the contract, the steady state falls back to the Nash equilibrium. The firm determines the feasibility of a payoff by calculating its net present value of discounted future payoffs.

Now, relaxing the Walsarian market assumptions, assume there are a fixed number of firms and a fixed entry condition (no fixed fee); however, there exists a fixed menu cost associated with changing existing workers' wages. For the firm, profit maximization occurs if the discount factor of the bargaining outcome is sufficiently low. Conversely, if a collusive 'Pareto' optimum strictly dominates the Nash equilibrium, it induces firms to revert to Pareto-inefficient Nash equilibrium. In a cooperative game, zero utility occurs when an incumbent threatens to revert from mark-up payoffs to Nash equilibrium, where the firm's marginal costs are equalized. Abbreviated punishment strategies make deterrence unprofitable.

### *5.1 Aggregate Technology Shock to the Economy*

When a median-scale aggregate technology shock affects the macroeconomy in the tn-1th period, this can be simulated by subtracting the change in output by a negative percentage for that given period. Effectively, this reduces the overall balance of the change in aggregate MRPL (the weighted productivity) in the tnth period, such that the aggregate decrease in output (a firm's gross revenue) is proportionate to the decreases in demand for goods and services as a result of the shock. The only way to offset this negative change and maximize profits for the new output is to either reduce the number of employed households (negative hiring) or reduce the labor market wage.

The second $t_{n+1}$ period of the simulation assumes that firms act logically and either reduce vacancies, reduce the nominal wage, or use a combination of both responses to find the point at which labor costs are low enough to optimize the MRPL during the $t_{n+1}$ period. However, households are keen to maintain the $t_{n-1}$ period wage $w$, which may be partially sustained by a group of firms but is not an optimal choice for the perfectly competitive firm to offer. The use of a negative hiring rate is a short-term tool for optimizing the MRPL and retaining the dominant bargaining position of firm $f$ with respect to the market wage $w$, which firms wish to offer households at every $t_{n+m}$ until the MRPL reaches *reservation* productivity, where $m$ is also the duration of the technology shock, after which unemployment equilibrium can be reached. At this point, $h$ approaches negative (signed) zero, and the labor market finds a new, lower equilibrium. $+h \in (0,1)$, $-h \in (-1,0)$ where $\beta$ ($e_u$) $\leftrightarrow$ ($e_m$).

*5.2 Changes to Wages and Hiring*

A firm can choose either to lower *w* or to reduce vacancies and engage in job destruction. How do these changes affect wages? As the various exogenous factors impose a limiting constraint on π, employed households are paid more than firms would like, that is, with respect to *w*. The unemployed households' Nash bargaining solution is improved at each interval of time *t*. However, firms wish to adjust this wage to the new equilibrium, which occurs at a lower wage, so they can maximize $\pi$ at this equilibrium. In other words, firms are operating inefficiently. Households are indeed hard done by, but such is the nature of, the movement along the curve in the model. It therefore follows that when providing social security, government programmes must be more responsive to aggregate changes in real wages.

Taking the model and changing the incentive to hire, as in the Pissarides/Mortensen model, n firms choose to fill job market vacancies if what we will refer to as each firm's 'weighted productivity' satisfies this inequality constraint, which is given by the composite function

$$\frac{h(x\alpha)t}{1+r} \; s.t. \;\; (x')\alpha \; \geq \; (x)t - n$$

where the constraint is also the labor market reservation productivity. Here, *x'*, the weighted productivity, is given by $x' = \pi/e_m$. Substituting for *x'* and *x*, the composite function becomes:

$$h\,(\pi^{\alpha}/em\,)/1 + r \quad s.\,t.\,(x')^{\alpha} \geq (\sum\pi/\sum em)^{t-n}$$

The above equation represents the equilibrium job creation conditions; what implies is that the marginal value of each firm's wage bargaining decision must be equal to or greater than the present value of their reservation productivities over the duration of the technology shock *t*.

*5.3 New Equilibrium Conditions for Wages and Prices*

The three-period dynamic model consists of $N$ downstream firms. Market share is equally distributed among these firms; hence, there are benefits to colluding, which are equal to $Nj + Ni = Nj{+}i$, where $j = \overline{\overline{N}} - j$ and the socially optimal merger is given by the total market share less merged market share. Downstream consumers bear the transportation cost of R for accessing the services of a competitor. If $R* < \overline{\overline{R}}(N)$ is the transportation cost of the new merged entity and is less than the price of the previously acquired firms, then consumers divert consumption to $N - j$ firms. In addition to the transportation cost, a *constant* menu cost *T* is levied between each competitor's bundles of goods. Output $Q$ is monotonic in accordance with the equilibrium price, *P*; thus, $\pi = p(q) > P'(Q')$, which is a markup set by each dominant firm. Additionally, there are $M$ potential entrants, who, by undercutting the market equilibrium, can profitably service the downstream market at each given period, paying an entry fee *E*. Firm $i$ will produce at a cost of $C1(Q1)$, where the profit function for firm *i* can be written as:

The three-stage model is as follows:

5.3.1 Period 1 – *P* set to maximize profit function

5.3.2 Period 2 – Decrease in *P* via Limit Pricing strategy (*The Nash Reversion*)

5.3.3 Period 3 – Increase in *P* to Regard any Loss

*5.4 Profit Maximizing Location Games*

In a downstream market, a competition authority is responsible for ensuring that the entrant is able to achieve a minimal profit, which can be given by the following function:

$$E(R) \; = \; Q[qPe \; - \; Ce] \; + \; E \; > \; 0$$

The above equation is an entrant's reaction function and is a profitable pricing condition for an entrant in our model. If *P* is significantly high, the function converges to 0, and entry is prevented. Thus, optimal spatial collusion (Huck *et al.*, 2003) on a unit circle is profitable if the revenue of a coalition of size *M* is strictly greater than the expected revenue $E(R)$ from a noncooperative equilibrium. It is possible to show that the result of the merged firm's cooperation reduces the consumer transportation cost $n$ by the distance *D* between the new firm and its two closest competitors. The termination fee levied by the joint incumbent on other competitors implies higher price tariffs for consumers who communicate with other competitors. Thus, the logical incentive for the marginal consumer or for those who wish to maximize consumer surplus (if prices are left unchanged) is to remain with their operator. Free-entry equilibrium for a market that is large and dominated by several large firms would typically yield zero profit (Eaton and Wooders, 1985).

*5.5 Conditions for Wage Growth*

The matching function (see Pissarides, 2000; Shimer, 2005), taking the Cobb–Douglas form and characterized by constant returns to scale $Y = X + I + K^{\alpha} L^{(1-\alpha)} + A$, where (Y) represents output, (X) represents total exports, (I) represents total imports, (K) represents capital, (L) represents labor, and (A) represents a mystery variable, quantified as the autonomous technology stock, research & development, which provides a succinct representation of the equilibrium as a function where the number of households looking for jobs and the proportion of capital employed by firms are strictly increasing in (Y).

Take the basic aggregate output equation in Cobb–Douglas form:

(5.5.1) $$Y(t) = K(t)^{\alpha} + L(t)^{1-\alpha} + A(t)$$

$$For\ al0 < \alpha < 1$$

Having stated the economy's production function. We derive the first-order conditions:

(5.5.2) $$\frac{\partial f}{\partial Y(t)} = \frac{\partial f^{\alpha}}{\partial K(t)} + \frac{\partial f^{1-\alpha}}{\partial L(t)} + \frac{\partial f}{\partial A(t)}$$

$$= \frac{\partial f}{\partial K(t)}(\sum\nolimits_{t}^{T} K)^{\alpha} + \frac{\partial f}{\partial L(t)}(\sum\nolimits_{t}^{T} L)^{1-\alpha} + \frac{\partial f}{\partial A(t)}(\sum\nolimits_{t}^{T} \dot{A})$$

By multiplying both variables on the right by output Y and plugging in the equation for balanced wage growth, we have the following convex expression for output, which is increasing in Y:

(5.5.3)

$$\frac{\partial f}{\partial K}\sum\nolimits_{t}^{T} K \frac{K(t)K(t)^{\alpha}}{Y\ K(t)} + \frac{\partial f}{\partial L}\sum\nolimits_{t}^{T} L \frac{L(t)L(t)^{1-\alpha}}{Y\ L(t)} + \frac{\partial f}{\partial A}\sum\nolimits_{t}^{T} A \frac{A(t)A(t)}{Y\ A(t)}$$

Next, a closed-form expression to express a Nash bargaining solution is generated where households optimize wage maximization decisions and firms choose profit maximization outputs. Here, if real wages are high, firm output is low, and vice versa:

(5.5.4) $$U = (n, w, p, q)$$

$$where\ n\ and\ q\ are\ i.i.d\ constants\ 0 < x < 1$$

In forming the following Hamiltonian, aggregate output can be subjected to the following constraint faced by the perfectly competitive firm:

(5.5.5) $Hn, q = Kq + Ln + A - nw(1 - w) + q(1 - pq)$

The staggered aggregate growth rate of capital ($K$) and labor ($L$) at time $t$ can be accounted for by the variables $n$ & $q$:

(5.5.6) $$\partial Y(t)\ \partial q(t) = K + q(1 - p)$$

$$and$$

$$\partial Y(t)\ \partial n(t) = -1n + (1 - q)$$

The Euler equation is applied to the above expression:

(5.5.7) $$\frac{\partial \mathrm{Y}}{\partial \mathrm{n}} / \frac{\partial \mathrm{Y}}{\partial \mathrm{q}}$$

Everywhere along the $dY(t)/dn(t)$ slope, firms and households agree on a nominal wage such that the expression can be said to increase in $q(t)$. The change in the discounted present value of future output maximization also changes the condition for optimal capital inputs at time $t$:

(5.5.8) $$K(t-1) + (1 - p(t))$$

$$= -1n(t) + (1 - q(t))$$

$$where\ K > 0\ and -1 > n > 1$$

Finally, equation 5.5.8 is simplified and rearranged to express the variable $K$ and then substituted back into equation 5.5.3 with a Taylor approximation:

(5.5.9) $$= \frac{\partial f}{\partial \mathrm{K(t)}}(\textstyle\sum_{t}^{T} -1n(t) + (1 - q(t))^{\alpha}$$

$$+ \frac{\partial f}{\partial \mathrm{A(t)}}\left(\sum_{t}^{T} A\right) + R(t)$$

$$+ \frac{\partial f}{\partial \mathrm{A(t)}}(\textstyle\sum_{t}^{T} A) + R(t)$$

The model takes into consideration the 'value of employment' and the trade-off of a switch to unemployment. Now, consider the value of employment for a short period, receiving a nominal wage. The probability of remaining employed for *t* periods is $e^{-r}(b)$.

(5.5.10) $$\int_{0}^{T} e^{-r(b)}(\textstyle\sum_{t}^{T} -1n(t))$$

$$+(1 - q(t))^{\alpha} + (\textstyle\sum_{t}^{T} L1 - p(t))^{1-\alpha} + \frac{1-e}{r+b} + \mathrm{R}(t)$$

*5.6 Other Reversion Strategies: Grim Reaper*

Now, consider a two-stage oligopoly game where the removal of any barrier to entry creates a nonobligatory fixed fee (*F*), which carries principal interest (*r*). Suppose the incumbent plays $x$–314 + $\beta\mu 2$ in time period 1. However, an entrant with a strictly dominant strategy is able to derive substantial profit by setting the price to derive a Pareto efficient payoff, e.g., $\geq x-304 + \beta\mu 2$, which dominates the games' previous Nash equilibrium. If the discount factor is high, profit maximization will occur, but this requires both the incumbent and the entrant to agree to a pure strategy to sustain a strictly dominated payoff. Any deviation from the Pareto-efficient Nash equilibrium is a nonprofit maximizing strategy.

Importantly, in the interest of maximizing consumer welfare, if there are avenues for galvanizing demand (in the form of public but noisy signals), a grim-trigger strategy can be maintained by both firms, with minimal secretive deviation (Aoyagi, 2002). The Pareto-efficient payoff is split evenly in the new collusive equilibrium, with the following outcome.

*5.7 Other Punishment Strategies: Abreu-type*

One major problem with the solution to a Pareto efficient collusive deviation is that while it may be optimal, a gradual increase in barriers to entry generally increases the number of firms at the free entry equilibrium (Harrington, 1991). An incisive tool for an incumbent in a co-operative Nash equilibrium is a punishment strategy that is determined by the avoidance of retaliation, where the entrant reverts. To a large extent, such strategies focus on two main forms of punishment: Abreu-type punishment strategies (Abreu, 1988) and Nash reversion (Friedman, 1971). Recent studies have demonstrated the effect of punishment strategies on the outcome of repeated games. (Wright, 2011). If, by simultaneously undercutting the incumbent firm, the competition concentration ratio increases, an entrant's Pareto-optimal profit function is given by profit from undercutting the incumbent's lower profit gained by colluding $\gamma - c$. If $\gamma > c$ for all $p^{\Omega} > p*$, then entry occurs if an entrant can control the competition created by $ei\, p^{\Omega} - pi$ via $n$ number of entrants through limit pricing strategies. However, as the $e^n$ number of entrants increases, perfectly competitive profits converge to zero.

Now, consider a similar market in which a specialized conglomerate operates in a single niche segment and the conglomerate owns several downstream subsidiary firms that are located within a designated geographical region, as with the Hotelling-Downs model of spatial collusion. The conglomerate must frequently allocate equity, resources and raw materials to each subsidiary, which collectively occupy a significant proportion of sales in the designated region. Assume that the market segment is a highly concentrated oligopoly of firms, where direct competition gradually emerges in the form of a number of discount rivals.

## 6. Discussion

*6.1 Labor Migration Controls*

As the Stark and Bloom (1985) labor migration model implies, employment is not the only determinant of life chances, and not everything can be explained by what an employment classification programme directly measures. However, for the social planner, point scoring is adequate for facilitating the mobility of workers who are sufficiently skilled but who are external laborers in relation to the domestic market. To explain the relevance of labor migration, points are given as a function of the finite form $0 < s < 1$. The variable $s$ is a mechanism for endogenously optimizing research and development or 'knowledge', although the actual variable is taken as given and exogenous. Each firm wishes to signal entry into a competitive market, hires a laborer into its domestic market and is required to use the system.

Firm (*f*) observes the productivity of household (a) who is employed and calculates a point score for ($H/a$) workers. Point scores are effective socioeconomic wage bands that act to categorize skilled individuals who are exceptionally qualified for remuneration within a particular sector. Two variables can internalize the concept of the point scores. The first is aggregate (a) productivity (i.e., qualifications/experience), and the second is (b) the nominal offered wage. We assume that all households have the ability to freely invest in improving their skills and productivity. In a more advanced model, point scores can be assigned to job vacancies as prerequisites that must be met by employers upon entry into the labor market. Assume that a potential worker (c), with a point score (s*), wishes to enter the market. Assume that household (b) also has a point score of (s*). A new worker (c) can enter the market if vacancies with point scores $[0 < s < s^*]$ exist within the labor market. If not, the new worker remains structurally unemployed. Such is the importance of the labor migration scores.

*6.2 Other Policy Implications: Labor Mobility*

Optimal labor mobility systems must offer a criterion for adopting a point-scoring system but must be governed by the perfectly competitive firm (f), where the effects are decentralized and the onus is placed on the perfectly competitive firm to negotiate and maintain balanced wage growth in the short run. As suggested previously, each firm must allocate a socioeconomic classification to each matched job vacancy. Households are then allocated 'mobility' point scores based on their skills. For instance, points between $s0$ and $s1$ equate to a specific socioeconomic classification. The vacancy bands are then matched with the mobility points. All the workers that have joined the workforce in the last time period subsequently become incumbents, who can supply their labor without the criterion of a 'mobility' point score. Therefore, in essence, all households and workers within the labor market will be of adequate productivity, given the aggregate point score level.

As suggested previously, this is the exogenous framework within which the real wage is derived. The underlying issue here is that, as stated in the introduction, when the labor supply is not fully mobile across markets, real wage rigidities exist. Specifically, the less mobility the labor market affords to workers in terms of point scores, the higher the nominal wage will be.

The solution to this is that, if we propose that firms be allowed to designate point scores, this will prevent direct control of industry-wide labor mobility vis-à-vis. This is opposed to the less appealing centrally controlled system where labor market mobility can be restricted or encouraged accordingly by the government, which causes wages to inflate and/or deflate, as explained by the model.

Price increases are an altogether overlooked paradigm concerning catalysts for labor mobility in the short run. While price increases are a reason for increasing the real wage, a high real wage is not itself a reason for increasing labor mobility. However, the real wage is low. Thus, when the price is high and the real wage is low, there is an argument for increasing the mobility of labor. As wage growth inflates over time, the social planner is tasked with optimizing both social surplus and producer surplus by maintaining a Pareto-efficient wage level. Real household income can be controlled for using central government legislation. One policy tool that is at the social planner's discretion is the extension of job protection for existing workers. What are the exact details of job protection laws? These include legislation aimed at reducing the job destruction rate. In other words, while wages are high, efforts can be made to ensure that firms do not overutilize the counter mechanism for reducing labor costs. For example, firms can be discouraged from terminating contracts for existing workers who have worked for a period of $t^{1+n}$ periods, where $t - n > 0$. The outcome of such legislation is simple. The existing wage equilibrium is prevented from becoming sticky, as firms cannot reduce employees who are hired or, as such, cannot renegotiate lower wages with new workers $e^n$.

Finally, what exactly comes to pass in those less fortunate economies when all but the lower echelon (based on socio-economic classification) of skilled labor migrate to more developed economies? What needs to be done in order to plug such a gap? This paper makes no attempt to convince the audience of its gravitas in understanding such a dilemma, however, we must necssarily think of the most compelling solutions to ease the burden of the model, should such a scenario come to pass. Efforts need to be made where possible to increase ubiquitous access to further and higher education for women in developing economies. This is at the heart of any such resolution. Why? Specifically, skilled laborers in industries such as healthcare, pharmaceuticals and medicine are highly transferable and can leave large cyclical gaps in employment once acquired. Where necessary, agriculturally inspired education, which involves understanding farming and development but in the primary sector, must be prioritized. These industries are systemically important and, as such, pose no threat to the developed labor market; however, because of the preindustrialized nature of capital and technology in such economies, there is no value in investing in unsustainable capital markets prior to any establishment of a bottom-up growth model where all workers can viably contribute to the production possibilities of such economies.

## 7. Conclusion

Using a formula for the extensive form outcome for an oligopoly seeking to protect cartel profits from entry, this paper has attempted to solve a similar case for a conglomerate firm. The social planner must always give due notice to the underlying intent of the pricing strategy of the firm. It has been shown that price reductions in spatial location games can affect the transportation costs of reaching the service of a newly merged firm when the price is strictly less than the cost of reaching either of the premerged firms, an effect that can encourage market participants to pursue cooperative '*code-sharing'* agreements.

The price effect of a coalition reduces consumer welfare by a proportional amount because, by undercutting an incumbent's 1st period price, the change in a generic consumer's marginal surplus over time must be strictly negative. Taking this further, there is also a negative limit price effect experienced by potential entrants, as the equilibrium plus the entrance fee exceeds marginal revenue. This effect can be shown by demonstrating that a merged firm can simply impose a higher fixed entry cost on potential entrants, which is greater than the cost of an outside option, i.e., higher than the revenue gained from participating in the market.